\documentclass{article}

\usepackage{arxiv}

\usepackage[utf8]{inputenc} % allow utf-8 input
\usepackage[T1]{fontenc}    % use 8-bit T1 fonts
\usepackage{hyperref}       % hyperlinks
\usepackage{url}            % simple URL typesetting
\usepackage{booktabs}       % professional-quality tables
\usepackage{amsfonts}       % blackboard math symbols
\usepackage{nicefrac}       % compact symbols for 1/2, etc.
\usepackage{microtype}      % microtypography
\usepackage{lipsum}		% Can be removed after putting your text content
\usepackage{graphicx}
\usepackage[round]{natbib}
\usepackage{doi}
\usepackage{subcaption}
\usepackage{graphicx}
\usepackage{xcolor}

\title{Exploring Drug Safety Through Knowledge Graphs: \\ Protein Kinase Inhibitors as a Case Study}

\author{
	\href{https://orcid.org/0009-0004-1731-0117}{\includegraphics[scale=0.06]{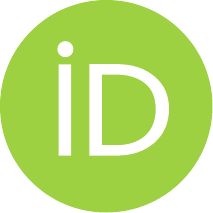}\hspace{1mm}David Jackson} \\
	Institute of Informatics\\
	University of Amsterdam\\
	Amsterdam, Netherlands \\
	\texttt{d.i.jackson@uva.nl} \\
	\And
	\href{https://orcid.org/0000-0003-4530-6110}{\includegraphics[scale=0.06]{orcid.pdf}\hspace{1mm}Michael Gertz} \\
	Institute of Computer Science\\
	Heidelberg University\\
	Heidelberg, Germany \\
	\texttt{gertz@informatik.uni-heidelberg.de} \\
	\And
    \href{https://orcid.org/0000-0002-4001-1164}{\includegraphics[scale=0.06]{orcid.pdf}\hspace{1mm}Jürgen Hesser} \\
	Data Analysis and Modeling in Medicine\\
	Mannheim Institute for Intelligent Systems in Medicine (MIISM)\\
	Medical Faculty Mannheim, Heidelberg University\\
	Mannheim, Germany \\
	\texttt{juergen.hesser@medma.uni-heidelberg.de} \\
}
\renewcommand{\shorttitle}{Exploring Drug Safety Through Knowledge Graphs}

\hypersetup{
pdftitle={A template for the arxiv style},
pdfsubject={q-bio.NC, q-bio.QM},
pdfauthor={David S.~Hippocampus, Elias D.~Striatum},
pdfkeywords={First keyword, Second keyword, More},
}

\begin{document}
\maketitle

\begin{abstract}
Adverse Drug Reactions (ADRs) are a leading cause of morbidity and mortality. Existing prediction methods rely mainly on chemical similarity, machine learning on structured databases, or isolated target profiles, but often fail to integrate heterogeneous, partly unstructured evidence effectively.
We present a knowledge graph-based framework that unifies diverse sources, drug-target data (ChEMBL), clinical trial literature (PubMed), trial metadata (ClinicalTrials.gov), and post-marketing safety reports (FAERS) into a single evidence-weighted bipartite network of drugs and medical conditions. Applied to 400 protein kinase inhibitors, the resulting network enables contextual comparison of efficacy (HR, PFS, OS), phenotypic and target similarity, and ADR prediction via target-to-adverse-event correlations. A non-small cell lung cancer case study correctly highlights established and candidate drugs, target communities (ERbB, ALK, VEGF), and tolerability differences.
Designed as an orthogonal, extensible analysis and search tool rather than a replacement for current models, the framework excels at revealing complex patterns, supporting hypothesis generation, and enhancing pharmacovigilance. Code and data are publicly available at \url{https://github.com/davidjackson99/PKI_KG}.
\end{abstract}

% keywords can be removed
\keywords{Adverse Drug Reactions \and Knowledge Graph \and Semantic Web \and Pharmacovigilance \and Drug Safety \and Network Analysis \and Protein Kinase Inhibitors \and Drug-Target Interactions \and Clinical Trials}

\section{Introduction}
\label{intro}
The World Health Organization (WHO) defines Adverse Drug Reactions (ADRs) as harmful reactions of a medicament \cite{edwards2000adverse}. Approximately 100,000 deaths per year are expected in the US alone \cite{ernst2001drug} and are hence of major concern. Different strategies have been developed for predicting ADRs so far. Among them are very recent machine learning approaches that use information about chemical structures and ADR-related databases \cite{dey2018predicting, mohsen2021deep, timilsina2022machine}. As with many machine learning approaches, a sufficient amount of training data is necessary to achieve high quality predictions. Especially for rare events that are reported, e.g., in the FDA Adverse Event Reporting System (FAERS) \cite{faers2024}, this is not guaranteed; and further on, one sees only the reported cases. Interestingly social media plays an important role in this context, too \cite{li2020combining} but this information is much more diffuse, sporadic, and unstructured. To associate these findings, however, with the diverse information sources on the internet is thus a challenge \cite{park2022large, daluwatte2020predicting}. Semantic web technologies provide a potential approach to collect and interpret the necessary information \cite{jiang2015mining}, and knowledge graphs (KGs), in particular, turn out to be an effective tool \cite{li2023visualization, zhang2021prediction, joshi2022knowledge}. A key ingredient also seems to be a strict orientation towards contextual ontology \cite{zheng2023personalized}. 

This paper demonstrates how far semantic web representations allow for an additional strategy to represent diverse and unstructured data sources on the internet in order to provide relevant information regarding adverse drug reactions. This is seen in the light that there are a variety of efficient tools and databases that help identifying potential interactions from chemical structure information to combining existing (textual) reports and apply machine learning (typical examples are \cite{daluwatte2020predicting, abacha2015text}).

In contrast to prevailing state-of-the-art methodologies, our approach introduces an innovative dimension by integrating a multitude of drug information sources concerning both drug targets and structural data, alongside with data derived from clinical trials and case reports, into a KG-based model. A key feature of our method is the comparison of drug profiles based on target similarity within this model, aiding in the prediction of ADRs by correlating similar drugs and utilizing target information for more accurate predictions. 

This approach enables a more contextual analysis, facilitating the detection of intricate patterns and relationships potentially overlooked in more traditional linear or compartmentalized analyses. 
Our approach further broadens classical KGs by enriching data with extra attributes to deepen semantic understanding. This enhances our model's ability to integrate and analyze diverse data, offering predictive insights over mere descriptive connections. Our method excels in deep semantic analysis, uncovering nuanced correlations, especially in drug profile comparison and risk score integration, exceeding the usual scope of KG methodologies. Employing graph analysis in conjunction with statistical tools, our method offers a nuanced and comprehensive understanding of ADRs, providing a distinct perspective in contrast to the predominantly data-centric and algorithm-focused approaches that are characteristic of current advanced methodologies.

By leveraging the integration of diverse data sources along with advanced graph analysis and statistical tools, this paper demonstrates how the proposed methods provide a more nuanced and interconnected analysis of ADRs, potentially revealing complex patterns and relationships that traditional linear methodologies might miss. The ultimate objective is to enhance the understanding and prediction of ADRs, offering a significant advancement over current state-of-the-art methods in pharmacovigilance and drug safety research.

Following this section, we start with an examination of the input data, detailing the datasets utilized in our investigation (Section \ref{datseources}). This is followed by an exploration of general graph properties, highlighting the significant characteristics of our analytical models. The segment on drug prediction then delves into the methodologies and outcomes of our predictive analytics, complemented by a case example that demonstrates these approaches in a practical scenario (Section \ref{results}). After the results, the Section \ref{methods} provides a comprehensive overview of the experimental and analytical techniques employed. Lastly, Section \ref{discussion} summarizes our key findings and their implications.

\section{Data Sources and Information Extraction}
\label{datseources}

Before focusing on the results in detail, we briefly summarize the components of our underlying model. ChEMBL\footnote{\url{https://www.ebi.ac.uk/chembl/}} provides drug names, mechanisms, approval stages, and further requisite information on bioactive molecules. Since our model embodies the current scientific consensus regarding such compounds, text data describing clinical trials is extracted from PubMed \cite{pubmed2024} using the National Institutes of Health's (NIH) e-fetch utility. MetaMap \cite{aronson2001effective}, a program developed to map biomedical text to the UMLS Metathesaurus \cite{bodenreider2004unified}, identifies the medical conditions from the clinical trial titles. Then, RobotReviewer \cite{marshall2016robotreviewer} annotates this PubMed data by returning information on the trial's PICO characteristics (Population, Interventions/Comparators, and Outcomes) and assesses trials for potential biases using the Cochrane Risk of Bias tool. Moreover, papers are linked to ClinicalTrials.gov \cite{clinicaltrials2024} by their respective National Clinical Trial (NCT) ID to complement and evaluate some of the results. 

Finally, we distill drug safety profiles from case reports managed in the FDA Adverse Event Reporting System (FAERS) \cite{faers2024} database. These data are cleaned and the proportional reporting ratio (PRR) scores for each drug-adverse event pairing is calculated in order to discern true side effects.

\section{Results}
\label{results}
In this section, we demonstrate the potential of our proposed method to analyze, integrate, and leverage large amounts of data regarding different drugs to enhance the understanding and prediction of ADR. First, we introduce the general properties and structure of our network model. Then, we explore how these properties facilitate drug prediction. Finally, we illustrate the network's functionality in detail through a specific example.

\subsection{General Properties}

We take a proof-of-concept approach, attempting to accurately embody the proximity of a specific drug class, which can then be expanded to a plethora of different drug groups. Our methods were applied to 400 drugs of the protein kinase inhibitors (PKI) class of drugs, given the existence of extensively researched PKIs alongside those that lack any study. Using the previously described materials, a query for the drug names and synonyms resulted in approximately 8,000 PubMed papers (1,908 of these include a full text) and 1,031 different UMLS conditions. It was possible to connect 2,739 of the papers to ClinicalTrials.gov. We scan all reports from the year 2016 to 2022 for our drug set, with a total of 11,353 adverse events found for 223 PKIs. 

We transform the identified concepts, drugs and conditions, into a network $N_{DC}$, by linking them based on their co-occurrence within the analyzed papers. Each connection (edge) in the network is assigned a weight, reflecting the strength of the association between the drug and the condition, with detailed methodologies to be elaborated in Section \ref{methods}. The resulting network $N_{DC}$ is a graph that incorporates $|V|=1287$ nodes (1,031 condition nodes and 256 drug nodes), $|E|=4081$ edges, and five connected components. Only 256 drug nodes are included, as no literature exists for the remaining 144 drugs. Furthermore, apart from one large component with 1,263 nodes, the other four components all contain less than 15 nodes. As the elements of such connected components do not exist in the larger context of our network, we will focus the following examinations on the large component $N_{Giant}$, which will make analysis methods such as diameter possible. Some of $N_{Giant}$'s basic properties are displayed below:

\begin{figure}[h]
\centering
\begin{subfigure}[c]{0.45\textwidth}
    \centering
    \begin{tabular}{|c|c|}
    \hline
    \multicolumn{2}{|c|}{$N_{Giant}$} \\
    \hline
    $|E|$ & 4056 \\
    \hline
    $|V|$ & 1263 \\
    \hline
    Average degree $deg(N_{Giant})$ & 6.42 \\
    \hline
    Average weight & 2.74 \\
    \hline
    Density $dens(N_{Giant})$ & 0.005 \\
    \hline
    Diameter & 8 \\
    \hline
    Degree Coefficient & -0.357 \\
    \hline
    \end{tabular}
    \caption{$N_{Giant}$'s basic network properties.}
    \label{fig:network_table}
\end{subfigure}
\hspace{0.25cm}
\begin{subfigure}[c]{0.4\textwidth}
    \centering
    \includegraphics[width=\textwidth]{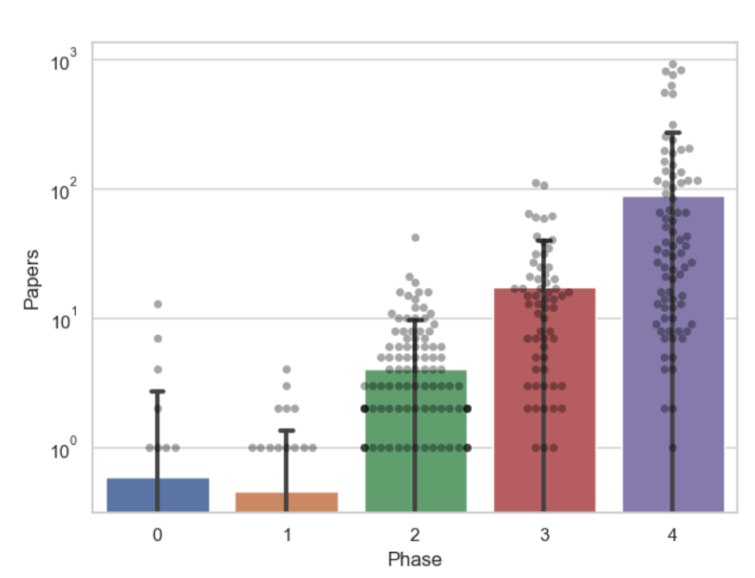}
    \caption{Number of papers for the PKIs grouped by their maximum clinical trial phase.}
    \label{fig:network_image}
\end{subfigure}
\caption{Overview of the $N_{Giant}$ component properties and research coverage of PKIs by clinical trial phase.}
\label{fig:combined}
\end{figure}

The vast majority of nodes with degree 1 are condition nodes, meaning there is a great number of rare conditions. This is underlined by the fact that the average degree of condition nodes is 3.95, whereas the average degree of drug nodes is 16.1. The node of the highest degree, 245, is 'Everolimus', and the condition node of the highest degree, 104, is 'Advanced Solid Tumor (C4329281)'. Table~\ref{tab:top_edges} lists the five edges with the highest weight in $N_{DC}$.

\begin{table}[h]
\centering
\caption{Top 5 most weighted edges in $N_{DC}$.}
\label{tab:top_edges}
\begin{tabular}{|l|l|r|}
\hline
Drug & Condition & Weight \\
\hline
Erlotinib & Non-small cell lung carcinoma (C0007131) & 273 \\
Gefitinib & Non-small cell lung carcinoma (C0007131) & 212 \\
Imatinib & Chronic myelogenous leukemia (C0023473) & 173 \\
Imatinib & Gastrointestinal stromal tumor (C0238198) & 126 \\
Sorafenib & Advanced Adult Hepatoma (C1706732) & 118 \\
\hline
\end{tabular}
\end{table}

\begin{figure}[h]
\centering
\includegraphics[width=\textwidth]{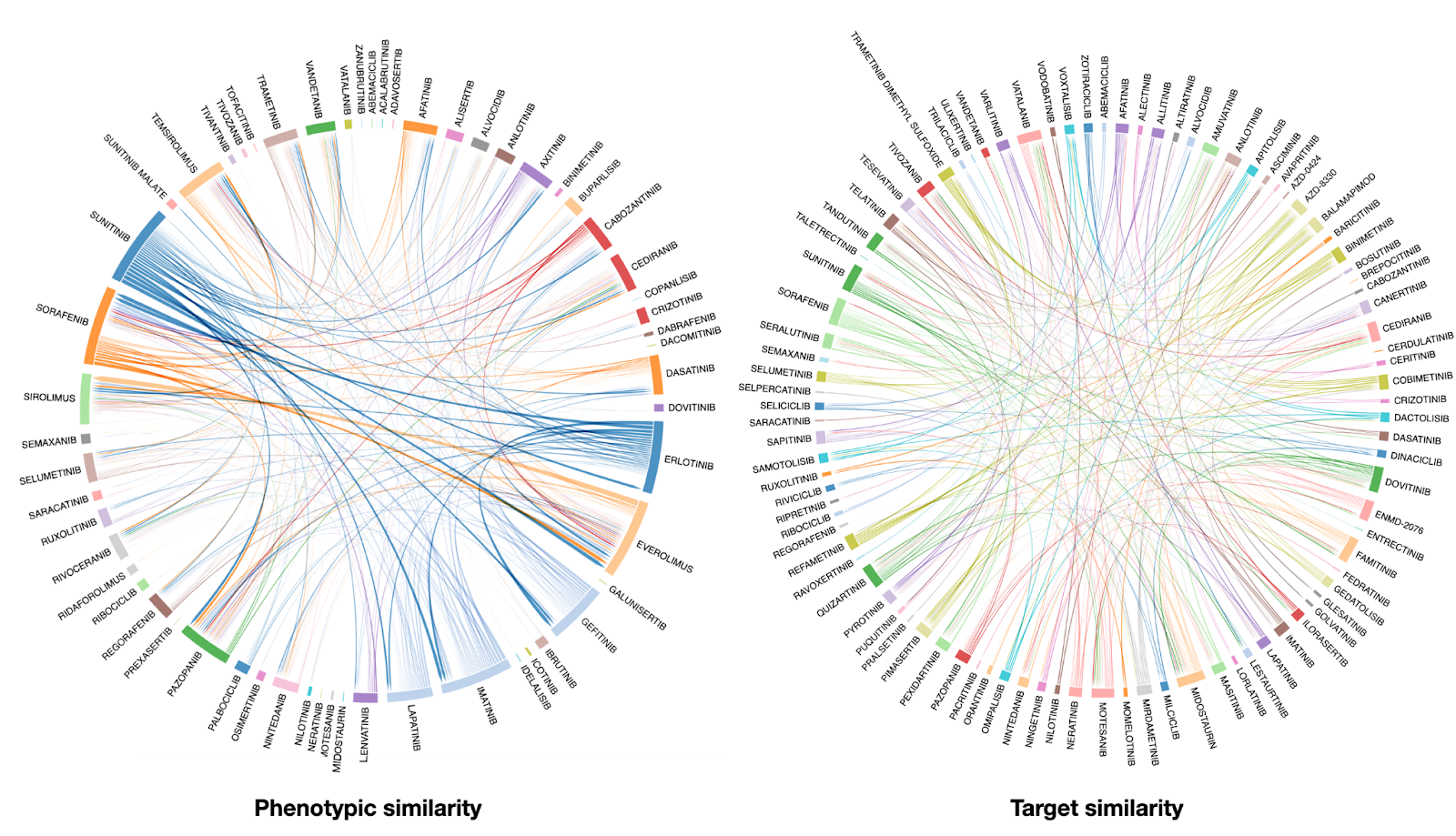}
\vspace{10pt}
\caption{Chord graph linking PKI pairings by the number of common related conditions (left) and chord graph linking PKI pairings by the number of common targets (right). A cutoff was introduced (left: > 10 conditions, right: > 1 targets and weight > 0.4) for clarity.}
\label{fig:chord_graphs}
\end{figure}

Each of the drugs shown in the table is approved for use for the condition it is associated with, proving that the simple edge weights (paper count) can help to draw conclusions regarding two concepts. As is obvious, cancer-related concepts are strongly represented in the dataset.
When two or more drugs are taken together, there is a possibility that they may interact with each other in a beneficial or harmful way. On the other hand, we must distinguish drug pairs that are tested or prescribed for the same reason but not together, i.e., in a study comparing Drug A with Drug B. $N_{DC}$ can reveal such similarities, which will help to find drugs of similar phenotypic effects. Figure~\ref{fig:chord_graphs} portrays these phenotypic analogies as modeled in the graph by linking drug node pairs by their combined neighborhood sets.
 
Note that the weights of $N_{DC}$'s edges are not taken into account, merely the overlap in conditions is examined. If the weights are included in this examination, less-known PKIs are not visible in the graph. Interestingly, we can detect some parallels between both chord graphs, as some of the PKIs that hold a large share of the phenotypic similarity graph such as Sorafenib, Sunitinib, or Vatalinib also possess several edges in the target similarity graph. By focusing on the most prominent associations shown in Figure~\ref{fig:chord_graphs}, it becomes clear that, naturally, there exists a bias for more widely known drugs such as Sirolimus and Everolimus. 

\begin{table}[htbp]
\centering
\begin{tabular}{|c|c|c||c|c|c|}
\hline
Drug $u$ & Drug $v$ & $|\Gamma(u) \cap \Gamma(v)|$ & Drug $u$ & Drug $v$ & $|\Gamma_w(u) \cap \Gamma_w(v)|$ \\
\hline
Sirolimus & Everolimus & 103 & Sirolimus & Everolimus & 281 \\
\hline
Sunitinib & Everolimus & 75 & Sorafenib & Everolimus & 260 \\
\hline
Sunitinib & Sorafenib & 74 & Sunitinib & Everolimus & 232 \\
\hline
Erlotinib & Gefitinib & 71 & Erlotinib & Gefitinib & 221 \\
\hline
Erlotinib & Sorafenib & 70 & Sorafenib & Sunitinib & 216 \\
\hline
\end{tabular}
\vspace{10pt}
\caption{Top 5 drug node pairs with most common neighbors in $N_{DC}$ where $\Gamma(u) $ denotes the set of neighbors of node u and $\Gamma_w(u)$ denotes the set of neighbors of node u including the weight between the endpoints.}
\label{tab:drug_comparison}
\end{table}

To adjust this imbalance, we apply a normalization function $S_{uv}$ to the similarity calculation:

$$S_{u,v} = \alpha \cdot \frac{|\Gamma(u) \cap \Gamma(v)|}{|\Gamma(u)| + |\Gamma(v)|} + (1-\alpha) \cdot \frac{|\Gamma(u) \cap \Gamma(v)|}{N},$$

where $\alpha$ is a parameter varying between $[0,1]$, and $N$ denotes the total number of nodes in the graph. Here, Sirolimus and Everolimus are, correctly so, still highly correlated but other less explored PKIs such as Brigatinib and Lorlatinib are also shown with a high score. $N_{DC}$'s bipartite nature enables the subdivision of drug and condition nodes. Thus, one can apply the same course of action on $N_{DC}$'s condition nodes to study what conditions seem to have similar treatments. 

\begin{table}[h]
\centering
\caption{Top 5 drug pairs with highest normalized similarity.}
\label{tab:normalized_similarity}
\begin{tabular}{|l|l|r|}
\hline
Drug $u$ & Drug $v$ & $\mathcal{S}_{uv}$ \\
\hline
Brigatinib & Lorlatinib & 0.184 \\
Sirolimus & Everolimus & 0.181 \\
Ripasudil Hydrochloride Dihydrate & Ripasudil & 0.165 \\
Itacitinib & Parsaclisib & 0.154 \\
Ulixertinib & Rabusertib & 0.154 \\
\hline
\end{tabular}
\end{table}

\begin{figure}[h]
\centering
\includegraphics[width=0.9\textwidth]{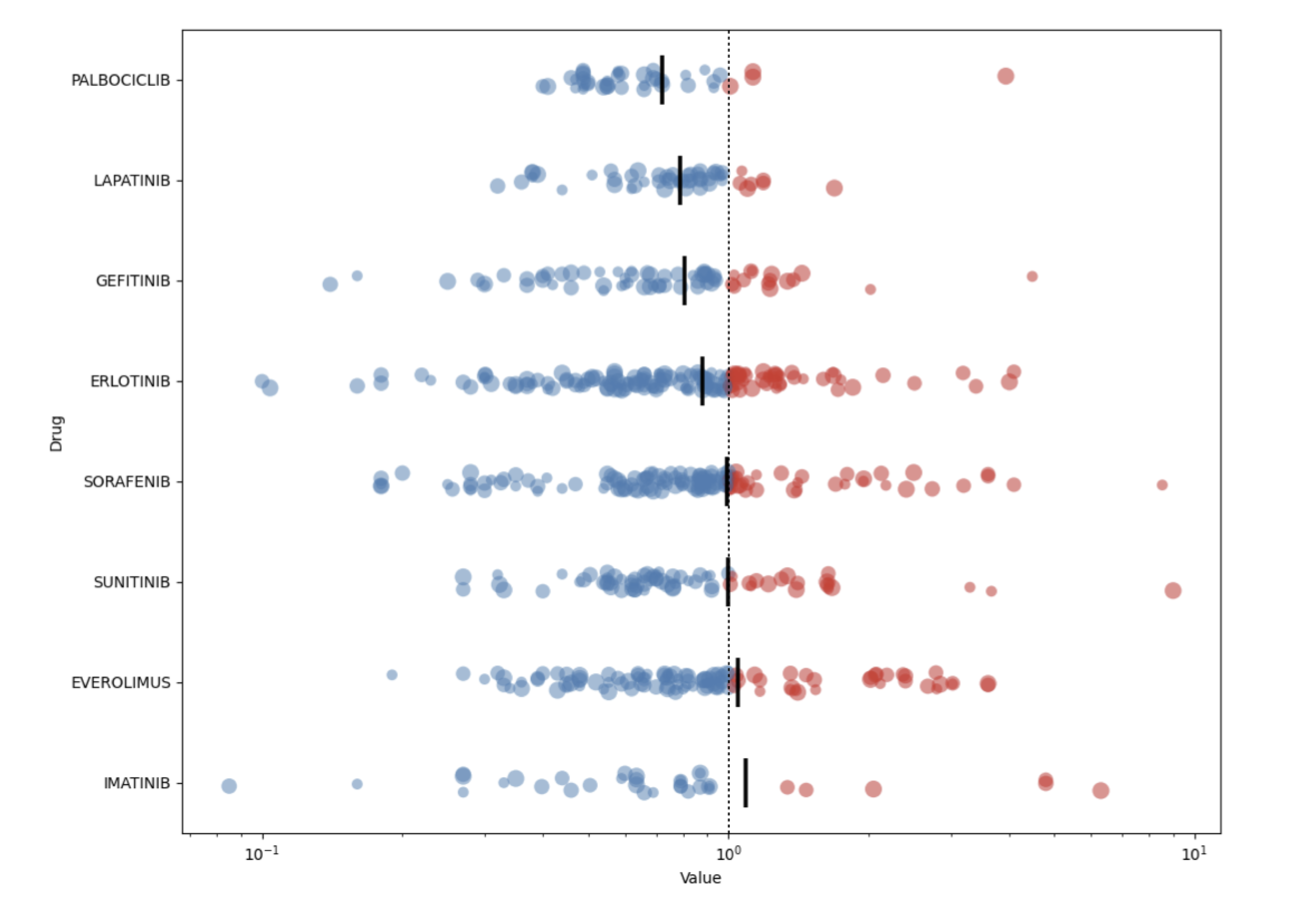}
\vspace{10pt}
\caption{Scatter plot displaying the HR scores collected from all papers for eight exemplary PKIs. Dots represent single papers and lines represent median HR values. Blue dots indicate a favorable effect and red dots represent a negative effect.}
\label{fig:hr_scatter}
\end{figure}

$N_{DC}$ allows for the comparison of values such as hazard ratio (HR), progression free survival (PFS), and overall survival (OS) not just along single relationships, but for the sum of relationships for one or multiple drugs. Figure~\ref{fig:hr_scatter} depicts such a comparison for 8 separate drugs.

Each of the given dots represents a paper regarding the individual drug and any possible disease. Furthermore, the papers may involve multiple drugs, diseases, or certain epidemiological facts, which makes this overview less conclusive. Imatinib, for instance, is one of the most widely used PKIs but it seems to have the highest mean HR. Nonetheless, such synopses help in understanding the current state of research surrounding individual PKIs.

\subsection{Drug Prediction}

Before identifying future drug prospects by comparing unapproved and approved drugs, it is crucial to assess if target/structural similarity implies phenotypic similarity. Target/phenotypic similarity is calculated for established drugs, considering both measured and curated target data, along with two types of phenotypic data (conditions and adverse events). Using three correlation measures results in a weak correlation (> 0.2) for most relationships. Despite weak correlations, they hold significance due to dataset discrepancies and complexities. Measured affinities show the least correlation with phenotypic effects, suggesting that similarities in measured affinities do not necessarily translate into phenotypic similarities. Drugs with high total target similarity weightings do have somewhat similar phenotypic effects, but the relationship is only slight. Identifying drugs based solely on composite target profiles is oversimplified and inaccurate. A more complex procedure is needed to predict a drug's phenotypic effects, requiring a deeper data analysis.

One approach is to focus on specific genes, phenotypes, or gene-phenotype relationships. For instance, gene 'B-RAF' is associated with 'Hand-foot Syndrome,' while 'ALK' is linked to 'lung-related diseases'. By correlating gene columns with the side effect column, one can assess gene-to-side effect relationships in the data. 
This method is applied to all genes in the dataset for which enough data exists. Table~\ref{tab:adverse_events} lists some of the predicted side effects for four paradigmatic targets.

\begin{table}[h]
\centering
\begin{tabular}{|l|p{8cm}|l|}
\hline
\textbf{Target} & \textbf{Adverse Events} & \textbf{Exemplary PKI} \\
\hline
B-RAF & Skin Reaction, Rash Maculo-Papular, Cholestasis, Hand-foot Syndrome, Blood Lactate Dehydrogenase Increase & Sorafenib \\
\hline
ABL & White Blood Cell Count Decreased, Cytopenia, Blood Test Abnormal, Fluid Retention, Pulmonary Edema & Imatinib \\
\hline
FKBP1A & Blood Cholesterol and Urea Increase, Proteinuria, Osteonecrosis, Interstitial Lung Disease, Mouth Ulcer, Aphthous Ulcer & Everolimus \\
\hline
VEGFR & Blood Pressure Increase, Hypertension Ageusia, Proteinuria, Glossodynia Flatulence, Oral Pain & Regorafenib \\
\hline
\end{tabular}
\vspace{10pt}
\caption{Four exemplary targets with the adverse events of highest correlation}
\label{tab:adverse_events}
\end{table}

One can quickly identify some confirmed target-to-adverse event relations, such as the previously mentioned connection between 'B-RAF' and 'Hand-foot Syndrome'.
The successful mapping of adverse events to targets by means of drug-to-adverse event correlations enables us to design a simple method for predicting any drug's adverse events given its target data. For a given drug, we obtain its adverse events by the calculated probabilities of all of its targets. 
One of the main drawbacks of this simple approach is the fact that genes that bind to more targets automatically will have more predicted side effects. However, no significant correlation (Partial Correlation Coefficient = 0.03) between the number of targets and the number of adverse events could be found.
Moreover, this simplified approach does not take the complex biochemical dependencies that may take part in target interactions into account. In other words, the co-occurrence of two targets may exacerbate a given phenotypic effect, whereas a target may also impede the effects of another target.				
We apply the same correlation method to predict drug indications and test it by comparing our results on non-approved drugs to possible drug indications given in the ChEMBL database. Drug indications in the ChEMBL database refer to the existence of early-phase trials examining the respective drug and disease. For non-approved drugs that contain possible indication entries, our predictions have an overlap of 46\% with the listed indications in ChEMBL.

\subsection{Case Example: Non-Small Cell Lung Cancer (NSCLC)}

After establishing the network's main characteristics and getting an overview of the underlying data, we can now further inspect the functionality of the network by performing an analysis on one specific example. Non-Small Cell Lung Cancer (NSCLC) is the disease of highest closeness centrality (excluding 'cancer' terms) and furthermore acts as an endpoint to the two highest weighted edges in the network $N_{DC}$.

\begin{table}[h]
\centering
\begin{tabular}{|l|c|c|c|c|c|c|c|l|}
\hline
\textbf{Drug} & \textbf{Weight} & \textbf{Effect} & \textbf{HR} & \textbf{PFS/OS} & \textbf{Approv.} & \textbf{AE} & \textbf{T} & \textbf{Receptor} \\
\hline
Erlotinib & 273 & \textcolor{green!60!black}{0.49} & \textcolor{green!60!black}{0.78} & 5.8/10.9 & 2004 & 43 & 132 & ERbB \\
\hline
Gefitinib & 212 & \textcolor{green!60!black}{0.50} & \textcolor{green!60!black}{0.68} & 8.5/15.9 & 2003 & 36 & 42 & ERbB \\
\hline
Osimertinib & 70 & \textcolor{green!60!black}{0.80} & \textcolor{green!60!black}{0.59} & 12.1/26.9 & 2015 & 46 & 7 & ERbB \\
\hline
Crizotinib & 52 & \textcolor{green!60!black}{0.66} & \textcolor{green!60!black}{0.44} & 11.4/33.9 & 2011 & 59 & 10 & EML4-ALK \\
\hline
Afatinib & 37 & \textcolor{green!60!black}{0.67} & \textcolor{green!60!black}{0.66} & 11.9/15.4 & 2013 & 53 & 22 & ERbB \\
\hline
Alectinib & 29 & \textcolor{green!60!black}{0.75} & \textcolor{green!60!black}{0.37} & 9.3/41.2 & 2015 & 58 & 5 & EML4-ALK \\
\hline
Sorafenib & 20 & \textcolor{green!60!black}{0.30} & \textcolor{green!60!black}{0.89} & 3.4/7.6 & (2005) & 42 & 131 & - \\
\hline
Vandetanib & 19 & \textcolor{green!60!black}{1.00} & \textcolor{green!60!black}{0.82} & 4.6/11.0 & (2011) & 73 & 20 & ERbB+VEGF \\
\hline
Ceritinib & 19 & \textcolor{green!60!black}{0.66} & \textcolor{green!60!black}{0.72} & 10.7/30.9 & 2014 & 67 & 10 & EML4-ALK \\
\hline
Anlotinib & 17 & \textcolor{green!60!black}{0.33} & \textcolor{red}{2.37} & 8.4/8.7 & - & 36 & 5 & VEGF \\
\hline
Brigatinib & 17 & \textcolor{green!60!black}{1.00} & \textcolor{green!60!black}{0.55} & 11.5/29.5 & 2017 & 49 & 3 & ERbB \\
\hline
\end{tabular}
\vspace{10pt}
\caption{Drug characteristics including weight, effect, hazard ratio, survival metrics, approval year, adverse events, and receptor targets}
\label{tab:drug_characteristics}
\end{table}

One must examine the given effect, HR, PFS, and OS scores with caution as an abundance of factors in the underlying literature can have a substantial effect on the outcomes, scattering the data. These factors could include study population differences, intervention differences, or paper biases. Obviously, the measures are less accurate for smaller-weighted drugs as outliers can have a higher influence. Comparing Erlotinib and Gefitinib, which will be done in greater detail later in this section, the mean HR, PFS, and OS values seem to be in the favor of Gefitinib. However, studies have shown that there is no significant difference between the effectiveness of Erlotinib and Gefitinib \cite{thomas2019comparative, lim2014comparison}. Nevertheless, studies have also shown Erlotinib to be less tolerable than Gefitinib \cite{thomas2019comparative, ma2013cost}, which is consistent with the results in Table~\ref{tab:drug_characteristics}.

\begin{figure}[h]
\centering
\includegraphics[width=0.8\textwidth]{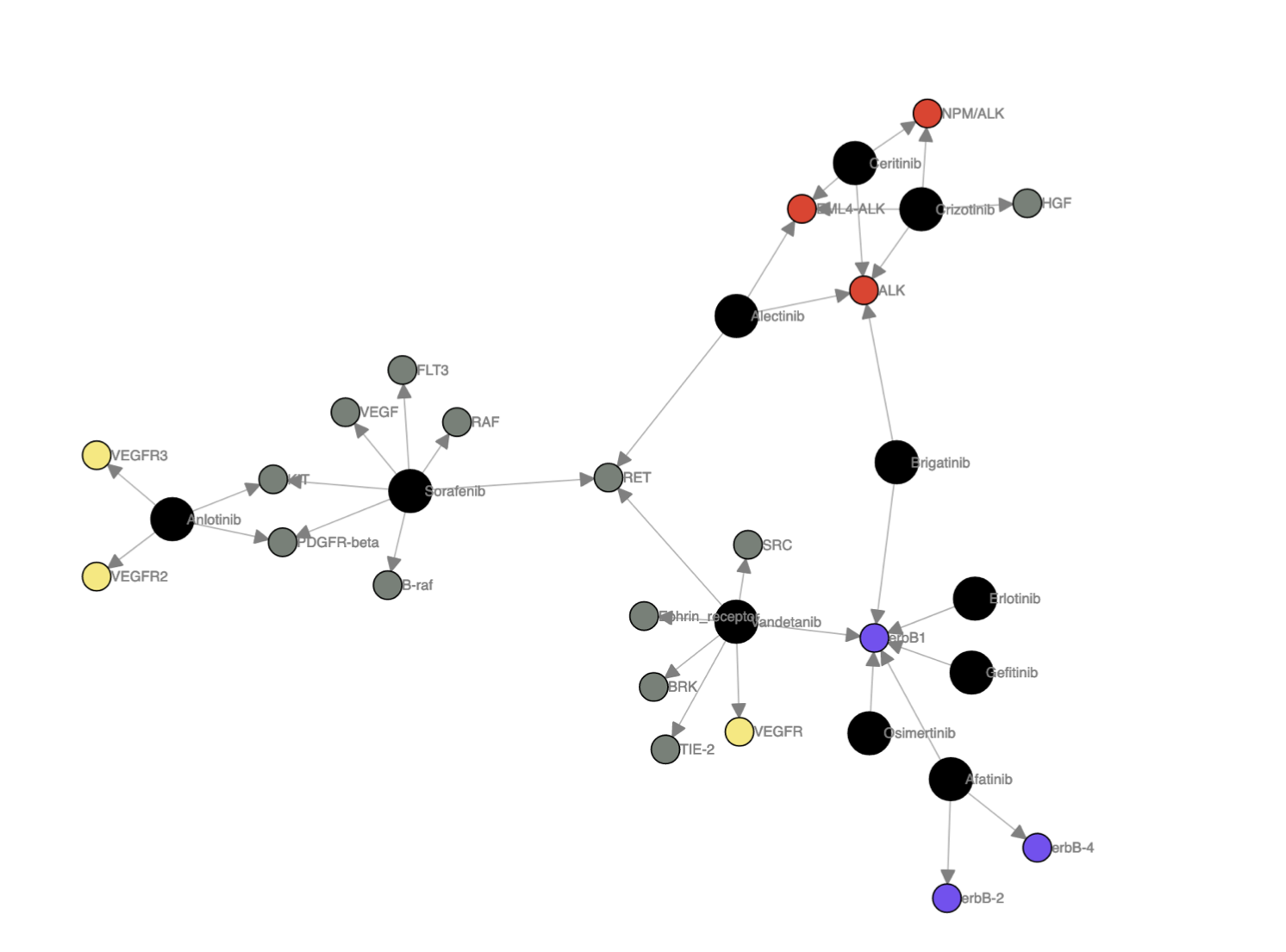}
\vspace{10pt}
\caption{Directed graph linking NSCLC drugs to the genes they bind to. Node colors are divided into drugs, VEGFR, ALK, ERbB, and other genes.}
\label{fig:target_graph1}
\end{figure}

While Vandetanib seems to have the highest amount of adverse events (PRR>3), one must keep in mind that this data might be influenced by the temporal aspect of the subset of reports (see Materials and Methods), which may also cause the two most popular NSCLC drugs, Erlotinib and Gefitinib, having fewer than average adverse events within the set. A striking observation is that all drugs (with two exceptions) bind to one of the three receptor classes ERbB, EML4-ALK, or VEGF. Of course, the drugs, in part, also bind to other receptors (see Figure~\ref{fig:target_graph1}, Sorafenib binds to 7 receptors) but this grouping was promptly noticeable. Coincidentally, one may observe that the only two exceptions, Sorafenib (which is primarily used for liver, kidney, and thyroid cancer) and Vandetanib (which binds to both ERbB and VEGF) are the only approved drugs not approved for NSCLC. Figure~\ref{fig:target_graph1} displays the manifestation of the complete target-to-drug connections from Table~\ref{tab:drug_characteristics}.

\begin{figure}[h]
\centering
\includegraphics[width=\textwidth]{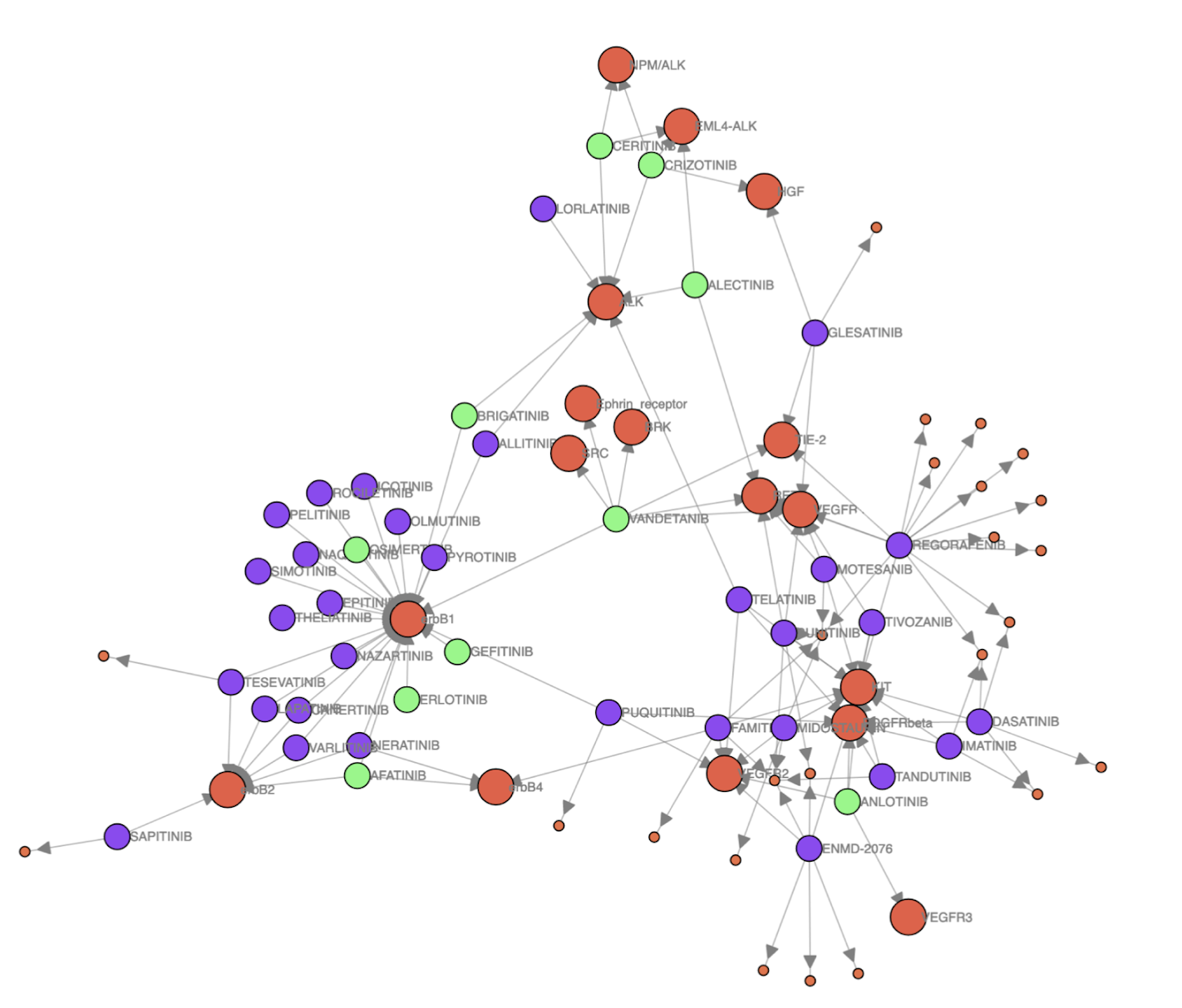}
\vspace{10pt}
\caption{Directed graph linking drugs to the genes they bind to, having two types of drugs and targets, respectively. Green nodes and large red nodes represent the drugs/targets from the original set in Figure~\ref{fig:target_graph1}. Purple nodes are the drugs with an overlapping target profile towards the drugs from the original set and small red nodes are the gene targets only hit by 'purple' drugs. Sorafenib was excluded here for clarity as it shares its targets with many other drugs.}
\label{fig:target_graph2}
\end{figure}

The node (gene) with the most incoming edges is 'erbB1' and the nodes (drugs) with the most outgoing edges are Sorafenib and Vandetanib. As previously stated, Sorafenib is the only drug not connected to a colored node. The target 'RET' seems to play a role in connecting the graph, which is broadly divided into three communities.			
Having constructed the structure and connectivity of the established NSCLC drug profiles, one can go a step further and incorporate additional drug nodes that have similar binding mechanisms. The purpose of this integration is to discover further drugs with prospective anti-NSCLC properties. Figure~\ref{fig:target_graph2} displays the result of this integration.

As can be seen, 'erbB2' still has the most ingoing edges. Erlotinib, Gefitinib, and Osimertinib share their curated target profile (only binding to ErbB1) with nine other PKIs, further increasing 'erbB1's in-degree. The only other drugs that share the exact curated target profile are Afatinib and Canertinib. Regorafenib has 15 gene targets. 12 drugs bind to KIT only one of which (Anlotinib) is from the starting set. Target 'erbB2', which is only targeted by one green node, now has ingoing edges by 6 additional drugs. The small red nodes mostly have degree 1 with a few exceptions on the bottom right of the visualization. No additional drugs bind to 'SRC', 'Ephrin receptor', and 'BRK' of the drug 'Vandetanib'.
One can further inspect Erlotinib, Gefitinib, and the drugs binding to ErbB1 by comparing their structural similarity to each other. Erlotinib and Gefitinib have a similarity of 0.35, while the other drugs of the highest similarity to both Erlotinib/Gefitinib are Icotinib (0.49/0.6) and Simotinib (0.32/0.25). Both Icotinib and Simotinib are non-approved drugs that were accurately predicted to have prospective anti-NSCLC properties as both drugs are in clinical trials for NSCLC. Canertinib (same target profile as Afatinib) is also a non-approved drug in clinical trials for NSCLC.				
Returning to the $N_{DC}$ network, one can apply the Adamic-Adar Index to some of the NSCLC-related drugs. The following tables result from this application.

\begin{table}[h]
\centering
\begin{minipage}{0.45\textwidth}
\centering
\begin{tabular}{|l|c|}
\hline
\textbf{Drug} & $A(x,y)$ \\
\hline
Erlotinib & 6.76 \\
\hline
Gefitinib & 5.98 \\
\hline
Sorafenib & 5.06 \\
\hline
Crizotinib & 4.98 \\
\hline
Afatinib & 4.41 \\
\hline
\end{tabular}
\vspace{10pt}
\caption{Top 5 recommended drugs for 'Osimertinib'}
\label{tab:osimertinib}
\end{minipage}
\hspace{0.25cm}
\begin{minipage}{0.45\textwidth}
\centering
\begin{tabular}{|l|c|}
\hline
\textbf{Drug} & $A(x,y)$ \\
\hline
Gefitinib & 35.75 \\
\hline
Sorafenib & 31.75 \\
\hline
Imatinib & 21.40 \\
\hline
Sunitinib & 21.28 \\
\hline
Everolimus & 20.74 \\
\hline
\end{tabular}
\vspace{10pt}
\caption{Top 5 recommended drugs for 'Erlotinib'}
\label{tab:erlotinib}
\end{minipage}
\end{table}

The recommendations for 'Osimertinib' strongly associate with the results from Table~\ref{tab:drug_characteristics}, whereas the recommendations for 'Erlotinib' include more well-known drugs. This is expected due to the uneven degree distribution in $N_{DC}$, where nodes of high degree will naturally get recommended more often, especially to other high-degree nodes. Lastly, the graph $N_{DC}$ also enables a closer comparison of Erlotinib and Gefitinib by means of a simple edge and punchline text query. Table~\ref{tab:clinical_trials} shows partial reviews of six clinical trials that directly compare the two drugs as saved in the network.

\begin{table}[htbp]
\centering
\caption{Partial review of six RCT's directly comparing Erlotinib and Gefitinib for the treatment of NSCLC.}
\label{tab:clinical_trials}
\begin{tabular}{|p{2cm}|p{7cm}|p{3cm}|c|c|}
\hline
\textbf{PMID} & \textbf{Punchline Text} & \textbf{Population} & \textbf{Part.} & \textbf{Bias} \\
\hline
27755796 & Median progression-free survival (PFS) was 30 months (gefitinib vs. erlotinib: 31 vs. 24 months; P > 0.05). & Patient \newline Drug Resistance \newline Japanese & 42 & 0.08 \\
\hline
22806307 & The CSF concentration and penetration rate of erlotinib were significantly higher than those of gefitinib (P = 0.0008 and <0.0001, respectively). & c-erbB-1 Genes \newline Cerebrospinal Fluid & 15 & 0.19 \\
\hline
21684626 & Exploratory analyses showed that there was no significant difference in RR and PFS in the gefitinib arm compared to the erlotinib arm (RR (\%) 47.9 vs. 39.6, p=0.269; median survival (months) 4.9 vs. 3.1, p=0.336). & Female \newline Smoker \newline Adenocarcinoma \newline Mutation & 96 & 0.09 \\
\hline
19680652 & There was an association between the disease control with gefitinib and erlotinib (p = 0.031). & Female \newline Non-Smokers \newline Adenocarcinoma & 21 & 0.06 \\
\hline
16467097 & After treatment with erlotinib or gefitinib, patients with EGFR exon 19 deletions had significantly longer median survival than patients with EGFR L858R (34 versus 8 months; log-rank P = 0.01). & Patient \newline EGFR Receptors & 70 & 0.12 \\
\hline
19765296 & Erlotinib was slightly superior to gefitinib in all measures of response, although the differences were not statistically significant. & Patient \newline Population & N/A & 0.07 \\
\hline
\end{tabular}
\end{table}

\section{Methods}
\label{methods}
This section delineates the methodologies employed in our study, detailing the data collection, preprocessing, summarization, graph integration, and analysis phases pivotal for investigating drug efficacy and safety profiles. It provides a thorough overview of the data sources and elucidates the techniques and tools used to collect, normalize, and analyze the vast datasets. The aim is to ensure the reproducibility of our findings and facilitate further research by offering insights into the construction and analysis of a complex network that bridges drugs with medical conditions based on empirical evidence.

The methodologies and analytical strategies detailed herein not only ensure that our results are interpretable but also guarantee reproducibility. For comprehensive transparency and to facilitate further research in this domain, the complete codebase of our framework, alongside the datasets and results, is made publicly available at \url{https://github.com/davidjackson99/PKI_KG}. In the following, we mention several computational tools that were used in this study.

\textbf{Data Collection}: The foundation of our analysis is built upon a comprehensive data collection phase, focusing on two main datasets: drugs, conditions, and research papers, alongside drug-related data such as target tissues and adverse events.

\textbf{Drugs, Conditions, and Papers}: Drugs were systematically retrieved from the ChEMBL database, yielding a dataset encompassing names, classifications, molecular properties, and drug indications. This initial step secured a total of 400 Protein Kinase Inhibitors (PKIs), laying the groundwork for subsequent network construction. Furthermore, the research phase of each drug was considered critical for understanding its development stage and research breadth. Conditions and their relations to drugs were extracted from the titles of research papers sourced from PubMed and PubMed Central, leveraging the NIH’s e-fetch utility. This process ensured that only randomized controlled trials (RCTs), pivotal for assessing drug efficacy, were included. Additionally, the extraction employed MetaMap to identify medical conditions from titles, refining our dataset to encompass relevant biomedical entities. Clinical trials information, augmenting the dataset, was obtained from ClinicalTrials.gov, providing structured details like the number of participants, intervention measures, and outcomes which are instrumental in evaluating the effect of drugs.

\textbf{Target Tissue and Adverse Events Data}: Target affinity data, essential for understanding drug-target interactions, were sourced from the Small Molecule Suite (SMS), enriching our dataset with selectivity values and binding assertions for over 400 PKIs against various targets. Adverse event data were compiled from the FDA's Adverse Event Reporting System (FAERS), offering insights into the safety profile of each drug by detailing patient demographics, drug information, and reaction data from case reports spanning 2016 to 2022.

\textbf{Data Pre-Processing}: Pre-processing involves normalizing and structuring the collected data to fit our model's requirements. This included standardizing drug names, resolving synonyms, and filtering relevant RCT articles based on predefined criteria such as publication type and content relevance to drug-condition efficacy.
For clinical trials and adverse events data, normalization procedures were applied to ensure uniformity across datasets, facilitating accurate comparisons and analyses.

\textbf{Data Summarization}: Data summarization aimed at condensing the extensive information into manageable, meaningful representations. This involved synthesizing key findings from RCTs, including therapeutic effects, bias estimations, population demographics, and outcome measures, into a structured format conducive to network integration. Adverse events data were similarly summarized, with a focus on quantifying drug-associated risks through the calculation of PRRs for each adverse event, enabling a systematic assessment of drug safety profiles.

\textbf{Graph Integration}: The culmination of our data collection and summarization efforts was the integration into a comprehensive graph. This graph represents a bipartite, directed, and weighted network, encapsulating drug-condition relationships derived from our datasets. Nodes represent drugs and conditions, while edges signify the evidence-based connections established through RCT analysis. The influence of each paper on our analysis is quantified by a unique weight for each paper, calculated using a specific formula. This weight is the sum of 1 plus the normalized values for the publication year, bias, quantity of data, and the number of citations each paper has. The normalization function scales each of these values to a range between 0 and 1. Further, the weight of a connection between a drug and a condition in our network is determined by an edge weight function. This function calculates the sum of the weights of all papers that support the relationship between the drug and the condition. 

\textbf{Graph Analysis}: The graph analysis phase focused on exploring the structural and relational dynamics within the network, employing techniques such as network visualization, centrality measures, and community detection to uncover patterns and insights into drug efficacy and potential adverse effects.

To realize each of the different methods, a number of pre-existing modules, which facilitate the implementation process, are utilized. The most key modules include:

\begin{itemize}
    \item chembl web resource client: ChEMBL's Python client for accessing ChEMBL data. \url{https://github.com/chembl/chembl_webresource_client} (Accessed: February 11, 2026)
    
    \item RRnlp: Provides access to RobotReviewer's key modules for extracting abstract information including effect, punchline text, bias, outcome, intervention, population terms, and more. \url{https://github.com/ijmarshall/robotreviewer} (Accessed: February 11, 2026)
    
    \item pyMetaMap: A Python wrapper for MetaMap which is used for extracting named entities. \url{https://github.com/AnthonyMRios/pymetamap} (Accessed: February 11, 2026)
    
    \item NIH's E-Fetch utility and requests: Tools for retrieving PubMed and PMC articles as XML files. \url{https://www.ncbi.nlm.nih.gov/books/NBK25499/} (Accessed: February 11, 2026)
    
    \item NetworkX: Used for network visualizations and graph analysis. \url{https://networkx.org/} (Accessed: February 11, 2026)
    
    \item spacy, re and nltk: Convenient string search with the help of regular expressions, which are used for pattern detections (such as identifying biomedical measures HR, PFS, OS) as well as simple NLP tasks such as sentence tokenization. \url{https://spacy.io/}, \url{https://www.nltk.org/} (Accessed: February 11, 2026)
    
    \item NumPy, scipy.stats: Mathematical tools for certain statistical problems. \url{https://numpy.org/}, \url{https://scipy.org/} (Accessed: February 11, 2026)
    
    \item D3Blocks: A Python module built on the graphics of D3 Javascript to visualize the data. \url{https://d3blocks.github.io/d3blocks/} (Accessed: February 11, 2026)
    
    \item pandas: For easy structuring and integration of the different data points. \url{https://pandas.pydata.org/} (Accessed: February 11, 2026)
    
    \item xml.etree, csv, json, pickle: Python modules for handling different data types from the individual resources. \url{https://docs.python.org/3/library/} (Accessed: February 11, 2026)
\end{itemize}

\section{Discussion}
\label{discussion}
The paper demonstrates how semantic web technologies can be used to collect information of adverse drug reactions from various data sources and how this can be distilled to a human understandable information representation. The role of the paper is a technology proof-of-concept rather than a comparison against other methods. In this respect it is orthogonal to existing technologies and can be easily combined with them in order to get a more detailed information collection and representation of adverse drug reactions. 

Our approach provides a significant advancement by enabling the integration of diverse datasets into a coherent model that facilitates better decision-making in clinical and pharmacovigilance settings. The compatibility of our system with existing technologies allows it to enhance, rather than replace, the current methodologies. This integration is crucial for evolving drug safety monitoring into a more comprehensive practice that leverages the full spectrum of available data.

\subsection{Potential and Limitations}

\textbf{As a Search Tool}: The KG-based model developed in this study is not only capable of aggregating complex data but also transforms it into actionable insights, making it an invaluable tool for healthcare professionals. This enhanced search capability supports rapid and informed decision-making, which is essential for effective treatment and safety monitoring.

\textbf{For ADR Prediction}: Our model's ability to predict adverse drug reactions before they occur has demonstrated potential, especially by linking genetic markers and specific ADRs. These predictions can be integrated into clinical practices to customize patient care plans, thereby reducing the likelihood of adverse reactions and improving treatment outcomes.

\textbf{Advancing Personalized Medicine}: One of the most promising applications of our research is its potential to drive the development of personalized medicine. By incorporating patient-specific data and genetic information into our analysis, we pave the way for treatments that are customized to the individual's genetic makeup, significantly enhancing the efficacy and safety of therapies.

\subsection{Future Directions}
The model's scalability and adaptability to include more varied data sources will be a focus of future research. Enhancing the accuracy and reliability of data integration will improve the model's utility and applicability in diverse clinical environments. Further research will also need to address the standardization of data inputs and validation techniques to minimize errors and maximize the model’s predictive capabilities.
In conclusion, this study has shown that semantic web technologies hold significant promise for improving the collection, integration, and application of data in the field of pharmacovigilance. By continuing to refine these technologies, we can enhance our understanding and management of drug safety, ultimately leading to better clinical outcomes and personalized treatment strategies.

\section{Related Work}
\label{related}
Several papers have previously attempted to automatically extract relations between biomedical entities using different approaches. For instance, the inspection of drug to disease relations: \cite{chen2015network} construct a drug-disease association network from confirmed drug-disease relations and apply a weighting method developed by \cite{zhou2007bipartite}. \cite{abacha2011hybrid} rely on two techniques, manual pattern detection and machine learning based on SVM classification, to extract the drug-disease relations from medical abstracts. \cite{jiang2022effective} make use of a large-scale molecular association network in combination with a graph embedding as well as a random forest model to predict drug-disease associations. \cite{gottlieb2011predict} leverage multiple drug-drug and disease-disease similarity measures for the prediction of novel drug-to-disease relations, also laying a focus on personalized medicine. Coinciding with the positive, curative effects of certain drugs are the often negative, adverse effects.

In \cite{galletti2022exploring} a T-ARDIS (Target-Adverse Reaction Database Integrated Search) database is created, which provides a direct link between proteins and adverse events and uses this connection to predict potential adverse events linked to proteins. A target-centric prediction method is defined that uses T-ARDIS information to train a combination of machine-learning classifiers to predict whether the modulation of a given protein is likely to result in adverse events. In contrast, deep learning is utilized by \cite{tomasev2021use} to create models that can continuously assess risk and predict adverse events based on data from electronic health records. A very different approach is taken by \cite{oconnor2014pharmacovigilance}, where a corpus of over 10.000 tweets is text mined for adverse drug reactions. In a related effort to integrate heterogeneous biomedical data using knowledge graphs, BASIL DB, a semantic web-based database that unifies sources such as ChEMBL, PubMed, and food composition databases, enables querying and prediction of health effects from bioactive compounds \cite{jackson2025basil}.

\bibliographystyle{plainnat}
\bibliography{references}  %%% Uncomment this line and comment out the ``thebibliography'' section below to use the external .bib file (using bibtex) .





% \end{thebibliography}

\section{Appendix}

\begin{table}[h]
\centering
\begin{tabular}{ll}
\hline
\textbf{Abbreviation} & \textbf{Meaning} \\
\hline
ADR & Adverse Drug Reaction \\
FAERS & FDA Adverse Event Reporting System \\
HR & Hazard Ratio \\
KG & Knowledge Graph \\
NCT & National Clinical Trial \\
NIH & National Institutes of Health \\
NSCLC & Non-Small Cell Lung Cancer \\
OS & Overall Survival \\
PFS & Progression Free Survival \\
PICO & Population, Interventions/Comparators, and Outcomes \\
PRR & Proportional Reporting Ratio \\
PKI & Protein Kinase Inhibitor \\
RCT & Randomized Controlled Trial \\
UMLS & Unified Medical Language System \\
\hline
\end{tabular}
\caption{List of abbreviations and their meanings}
\label{tab:abbreviations}
\end{table}

\end{document}